\documentclass[]{emulateapj}
\shorttitle{Comparison of MHD and Hydro turbulence}
\begin{document}
\title{Comparison of spectral slopes of Magnetohydrodynamic and
Hydrodynamic turbulence and measurements of alignment effects}
\author{A. Beresnyak, A. Lazarian}
\affil{Dept. of Astronomy, Univ. of Wisconsin, Madison, WI 53706}
\email{andrey, lazarian@astro.wisc.edu}
%\date{\today}

\def\L{{\Lambda}}
\def\l{{\lambda}}

\begin{abstract} 
  We performed a series of high-resolution (up to $1024^3$) direct
  numerical simulations of hydro and MHD turbulence. Our simulations correspond
  to the "strong" MHD turbulence regime that cannot be treated perturbatively. We found
  that for simulations with normal viscosity the slopes for energy spectra of
  MHD are similar to ones in hydro, although slightly more shallower.
  However, for simulations with hyper viscosity the slopes
  were very different, for instance, the slopes for hydro simulations
  showed pronounced and well-defined bottleneck effect, while the MHD
  slopes were relatively much less affected.  We believe that this is
  indicative of MHD strong turbulence being less local than Kolmogorov
  turbulence. This calls for revision of MHD strong turbulence models 
  that assume local ``as-in-hydro case'' cascading. Non-locality
  of MHD turbulence casts doubt on
  numerical determination of the slopes with currently available
  ($512^3$--$1024^3$) numerical resolutions, including simulations
  with normal viscosity.  We also measure various so-called alignment effects
  and discuss their influence on the turbulent cascade.
\end{abstract}


\section{Introduction}

Turbulence is ubiquitous in astrophysical fluids which are characterized
by high Reynolds numbers. It affects key astrophysical processes, e.g. star formation
(Elmegreen \& Scalo 2004, McKee \& Ostriker 2007). The observational
signatures of turbulence are numerous and well documented.
For instance, random fluctuations on all scales, which is a sign of turbulence, has
been detected by a variety of observational techniques (see Crovisier
\& Dickey 1983, O'Dell \& Castaneda, Armstrong et
al. 1994, Lazarian 2009).
Turbulence is universal, because the laminar flows
with high Reynolds numbers are practical impossibility.
It is driven by a variety of mechanisms such as supernova explosions
starbursts, stellar winds, AGN jets, etc.
In Keplerian flows turbulence is generated by magnetorotational instability (Velikhov 1959, Chadrasekhar 1960, Balbus \& Hawley 1991). Galactic disks are subject
to the cosmic ray-induced Parker's instability (Parker 1966). 

Although, historically, hydrodynamic turbulence has been applied
to astrophysics, now it is accepted that for almost all
astrophysical fluid flows are coupled with magnetic fields, at least on
large scales. This necessitates the use of
the dynamic equations that include electric currents and magnetic
fields. The simplest approach in this respect is the continuous
non-relativistic one-fluid description, known as magnetohydrodynamics or MHD. 
This approach is broadly applicable to most of
the astrophysical environments, such as Solar wind,
Interstellar and Intercluster Medium, molecular clouds, stars interiors,
and so on, although there are some exceptions, such as ultra-relativistic
jets and shocks, where full relativistic equations should be used or small
scales of Molecular Clouds, where, due to relatively low ionization rate, the
two-fluid description of ions and neutrals is more appropriate.

The importance of astrophysical turbulence inspires much of theoretical
and numerical work aimed at understanding its properties. We should clarify, however,
that there are different types of MHD turbulence.
In this paper we deal with strong MHD turbulence, which can not be treated
perturbatively. The theory of weak Alfv\'enic turbulence, which has limited
applicability to astrophysics, is discussed elsewhere
(Sridhar \& Goldreich 1994, Ng \& Bhattacharjee 1996,
Lazarian \& Vishniac 1999, Galtier et al. 2000, 2002). 
In order to study the basic properties
of MHD cascade and to be able to directly compare to previous work, we restricted
ourselves to so-called balanced MHD turbulence or turbulence with zero net cross-helicity.
The properties of {\it imbalanced turbulence} are studied in a companion
paper Beresnyak \& Lazarian (2009).

The issue of the spectral slopes of MHD turbulence has caused a substantial
interest recently. A sizable number of papers attempting to measure
true asymptotic spectral slope of high-Reynolds number MHD turbulence from
direct numerical simulations appeared to date. 
For example, simulations
of weakly compressible MHD turbulence performed in \citet{haugen2003, haugen2004}
used finite-difference code with numerical resolution of up to $1024^3$
with explicit viscous and resistive dissipation. The energy spectral slope for
MHD was observed to be shallower than $-5/3$ which was interpreted as
the influence of the bottleneck effect. 
Another example is the paper of
\citet{muller2005} who measured the spectral slopes
of decaying MHD turbulence without mean field
and driven MHD turbulence with strong mean field. The pseudospectral
method with ordinary viscosity was ran at a numerical resolution
of up to $1024^3$. The authors argued that the slope was close to $-3/2$
in the mean-field case.

The motivation behind these and many other papers was to understand the
nature of the turbulent cascade. The turbulent energy transfer is,
in a sense, a central issue of turbulence,
be it hydrodynamic or MHD. And while hydrodynamic turbulence has its
``Standard Model'', the nature of MHD cascade is still debated.

 An important first step in turbulence theory
was made by Iroshnikov (1963) and Kraichnan (1965) who noticed that there is
a local magnetic field which can not
be excluded by a choice of reference frame, like the average
velocity in hydrodynamics. Furthermore, they assumed that turbulence
is weak, because perturbations are smaller than the mean field.
This implicitly assumed local isotropic dynamics and
happened to be a mistake, since the turbulent cascade
preferred perpendicular direction, i.e. produced
perturbations that are more and more
anisotropic, this way increasing interaction and preventing
turbulence from becoming weak.
% In addition,
%the Iroshnikov-Kraichnan turbulence is isotropic, which was proved later to be also
%incorrect.

The understanding of various aspects of MHD turbulence,
including the role of turbulence anisotropy, compressibility,
etc. resulted in a number of publications (see Dobrowolny, Mangeney, \& Veltri 1980, Shebalin, Matthaeus \&
Montgomery 1983, Montgomery \& Turner 1984, Higdon 1984). For the
most part of the paper we will consider incompressible
MHD turbulence, which properties are dominated by the Alfv\'enic
perturbations. Interestingly enough, some properties of Alfv\'enic
turbulence carry over not only to nearly incompressible low Mach
number flows, but also for flows with Mach numbers larger than
unity (Cho \& Lazarian 2003, see also \S 9.3).
 
It has been realized that interactions of Alfv\'enic modes in MHD 
turbulence has a tendency of getting stronger as the cascades unfolds.
Goldreich \& Sridhar (1995, henceforth GS95)
proposed a particular model of strong turbulence when
the interaction strength is being controlled by two competing
processes: a perpendicular cascade (a concept rigorously developed
in a theory of weak Alfv\'enic turbulence, e.g. Galtier et al 2000),
which tends to increase the interaction, and a decorrelation due
to cascading which tends to increase the frequency of perturbations,
and thus decrease the interaction. GS95 concluded that the interaction
has to be marginally strong (``critical balance''), and therefore,
the cascade has to be of strong Kolmogorov type and has a spectral slope
of around $-5/3$. Predictions of this
model, such as scale-dependent anisotropy, were subsequently observed
in three-dimensional simulations (Cho \& Vishniac 2000,
Maron \& Goldreich 2001, etc).

Recently, however, a number of models, motivated by numerical spectral
slopes shallower than $-5/3$, appeared
(Boldyrev 2005, 2006, Gogoberidze 2007). The interest to this field
has been heated by the numerical discovery of so-called
scale-dependent polarization alignment (Beresnyak \& Lazarian 2006),
which was interpreted in 
a subsequent publication by Mason et al (2006) in favor of Boldyrev's (2006) modification
of GS95 model.

This paper is bringing attention to serious difficulties
that appear when one tries to measure true asymptotic slopes from
direct three-dimensional numerical simulations which have rather moderate Reynolds
numbers. By comparing spectra of hydrodynamic and MHD turbulence we
found that MHD turbulence might be much more nonlocal that Kolmogorov
turbulence, and, therefore, require much higher resolution to obtain
spectral slope by brute force approach. The second part of this paper
brings polarization alignment to more numerical scrutiny and compares
the simulation results to theory. Our simulations allow to test some of the existing conjectures about
properties of MHD turbulence. For instance, these results allow us to reject
a conjecture in \citet{boldyrev2006} that the alignment is limited
by the magnitude of local field wandering.

In what follows in \S2 we describe our approach based on comparing the properties of MHD and hydrodynamic turbulence,
present our numerical setup in \S3, present spectra in \S4, discuss anisotropy in \S5, discuss the interaction weakening and the alignment effects in \S6 and \S7, respectively, and provide some more hints of the non-locality of the MHD cascade in \S8. In \S9 we compare our numerical results with the existing theoretical predictions, as well as with the previous numerical work; we also discuss the applicability of findings obtained with incompressible simulations to the real-world compressible astrophysical turbulence. Our conclusions are summarized in \S10. 

\section{Slope measurements}

Various claims were made on the value of
MHD spectral slope, most of which were motivated by either Kolmogorov
$-5/3$ slope of strong turbulence (Kolmogorov 1941, GS95), or various versions
of $-3/2$ slope (Iroshnikov 1963, Kraichnan 1965, Gogoberidze 2007, Boldyrev 2006, etc).
Most numerical studies aimed to confirm either of the above.
A number of critical issues were overlooked, though. Below we present
a novel perspective to slopes measurements in MHD. It turns out that
it is incorrect to measure slopes directly from 3D numerics due to
a {\it systematic error} that comes with so-called bottleneck effect.
Also, it is impossible to measure MHD slope by comparison with hydro slope,
because the bottleneck effect turns out to be different in hydro and MHD.
This difference, however, allows us to criticize models that rely on
strong local Kolmogorov cascading, such as GS95 or Boldyrev (2006)
\footnote{Local cascading models normally use a formula $\epsilon=\rho v_l^2/\tau_l$,
where $v_l$ is the velocity perturbation on scale $l$, $\tau_l$ is the cascading timescale
and $\epsilon$ is a constant equal to the energy dissipation rate per unit volume.}.

%In this paper we compare spectral slopes expected in theoretical models with
%the numerical results.
To be precise,  $-5/3$ is not the exact predicted slope for 
incompressible hydrodynamics. This number comes from the Kolmogorov self-similar
cascade, but soon it was realized that realistic turbulence
is not exactly self-similar. To correct for this, various models
of {\it intermittency} were proposed (Kolmogorov 1962, Obukhov 1962),
with the most popular being She-Leveque model (1994) in which the predicted
slope is around $-1.70$. This slope was very close to what was observed
in highest-resolution DNS of \citet{kaneda}. In the aforementioned work
it was possible to separate spectrum into inertial range and relatively more
flat part that was due to a bottleneck effect.
Although modifications of S-L
model for MHD turbulence have been proposed, numerical studies
still often compare the spectrum slope with $-5/3$.
Similarly, when one proposes a model of turbulence with
$-3/2$ spectral slope it is often that
numerical studies aim to find an {\it exact} correspondence with this
slope, without regard for intermittency. We believe that this
is one important stumbling block in numerical determination of slopes.

The other, probably much more important misunderstanding, is to disregard
the {\it systematic error} that any numerical measurement of slopes
in DNS brings with it. The bottleneck effect is a pile-up
of energy before the dissipation scale due to the relative lack
of energy in the dissipation range \citep[see, e.g.,][]{falkovich}.
Due to relatively low resolution of currently available
simulations, this systematic error is always present.
To make it worse, most researchers present simulations with the
highest numerical resolution only. Although the amount of
numerical resources available to different groups differ
substantially, most ``high resolution'' simulations to-date
have numerical boxes between $512^3$ and $1024^3$. From the point
of length of inertial interval, and the influence of bottleneck
effect, the differences in linear scale of the multiple of two
are tiny. Another systematic error comes from the effect of
the driving scale. Often there is a dip right after
the driving scale, an anti-bottleneck effect of sorts, which
appears, possibly, due to the excess of energy on the driving
scale. We are not aware of driven turbulence simulations
that were able to get rid of this effect.
We belive that disregarding these two effects
and present numerical slopes as having
no systematic error at all is wrong.

In this paper we compare hydrodynamic
and MHD energy slopes obtained with the same code, the same driving
and exactly the same linear dissipation. Since there
are good theoretical predictions for asymptotic isotropic
hydro turbulence, we can try to use those. If one finds
that the nature of energy transfer in MHD and hydro is
similar (this is suggested in GS95 model where a strong
local Kolmogorov-like cascade is assumed), then we can directly
compare MHD and hydro slopes and make statements on MHD slope.
Unfortunately, as we show in the two subsequent sections,
this is not the case.
The defining feature of our simulations is the use of different
types of linear dissipation, namely natural viscosity and hyper-viscosity.
Although there had been some similarity in spectral slopes of MHD and hydro
in normal-viscous case, the hyper-viscous cases were very different.
This suggests that the nature of MHD and hydro cascades
are different and one can not use slope comparison between MHD
and hydro to get rid of the aforementioned systematic error.

\section{Numerical setup}

Incompressible MHD and Navier-Stokes equations can be written
in the following simple form

\begin{equation}
\partial_t{\bf w^\pm}+\hat S ({\bf w^\mp}\cdot\nabla){\bf w^\pm}=-\nu_n(-\nabla^2)^n{\bf w^\pm},
\end{equation}

where $\hat S$ is a solenoidal projection and ${\bf w^\pm}$ (Elsasser variables)
are defined in terms of velocity $\bf{v}$
and magnetic field in velocity units ${\bf b=B}/(4\pi \rho)^{1/2}$
as ${\bf w^+=v+b}$ and ${\bf w^-=v-b}$. Navier-Stokes equation is a
special case of equations (1), where $b\equiv 0$
and, therefore, both equations are equivalent with $w^+\equiv w^-$.
The RHS of this equation is a linear
dissipation term which is called viscosity or diffusivity for $n=1$ and
hyper-viscosity or hyper-diffusivity for $n>1$. Here we assumed that
viscosity and magnetic diffusivity is the same for velocity and
magnetic field. This is almost never true for realistic astrophysical
plasmas. However, as long as one wants to study the dynamics of large
scale turbulence, it is acceptable. This is because
the dissipation terms in today's numerical simulations
are never as small as to simulate {\it real} physical dissipation, instead,
they are used to remove energy on small scales and assure stability
of the code. In finite-difference codes a numerical
dissipation is always present and the linear dissipation terms
are often omitted altogether (see, e.g., Stone et al 1998).
In pseudospectral code, such as our own, the energy is conserved
with rather good precision, so the use of linear dissipation is necessary.

\begin{table}
%\large
  \begin{tabular}{c  c  c  c  c  c  c}
    \hline\hline

Run  & $n_x\cdot n_y\cdot n_z$ & x:y:z  & $B_0$ & $\Delta t^*$ & $f$ & dissip. \\
   \hline
   H1      &  $512^3$        &  1:1:1   & --   &   16  &  $v$      &  $4.5\cdot10^{-4}k^2$ \\
   H2     &  $768^3$        &  1:1:1   &  --  &   10  &  $v$      &  $7\cdot10^{-13}k^6$ \\
   H3     &  $1024^3$       &  1:1:1   &  --  &   7   &  $v$      &  $2.2\cdot10^{-13}k^6$ \\
   \hline
   M1   &  $512^3$        &  1:1:1   &  1   &   24  &  $v$      &  $4.5\cdot10^{-4}k^2$ \\
   M2  &  $768^3$        &  1:1:1   &  0   &   10  &  $v$      &  $7\cdot10^{-13}k^6$ \\
   M3  &  $768^3$        &  1:1:1   &  1   &   10  &  $v$      &  $7\cdot10^{-13}k^6$ \\
   M4 & $512\cdot768^2$ &  10:1:1   &  10  &   10  &  $v$      &  $7\cdot10^{-13}k^6$ \\
   M5 & $512\cdot1024^2$ & 10:1:1   &  10  &   10   &  $w^\pm$  &  $2.2\cdot10^{-13}k^6$ \\
   \hline
  \end{tabular}
  \caption{Description of simulations. * - $\Delta t$ is the duration of the high resolution runs, prior to that
 we ran lower resolution runs for a long time to ensure stationary state. E.g. we ran $B_0=0$ case for 180 time units prior to
going to high resolution. Only last 3-4 time units of high resolution runs were used for measurements. One time unit
corresponds to an eddy turnover time of the largest coherent eddies.}
  \label{experiments}
\end{table}

We evolved incompressible MHD and Navier-Stokes equations in time
using a well-known pseudospectral technique \citep[see, e.g.,][]{cho2000}.
We have chosen pseudospectral code as it allows precise control over dissipation.
%The incompressible MHD equations written in Fourier space were treated
%as ordinary differential equation with respect to time for each and
%every separate space Fourier harmonic. The nonlinear term contribution
%was calculated in real space and tranformed back into Fourier space.
%The incompressibility was preserved by applying a projection
%operator to the nonlinear term.
Our hydro code was identical to the MHD one, except, naturally,
for the lack of magnetic field.
The summary of high-resolution runs is presented in Table 1.

We performed four types of simulations: a simulation of statistically
isotropic hydro turbulence; a simulation of well-developed stationary
MHD turbulence without mean field ($B_0=0$), which also has been called
statistically isotropic MHD turbulence; a simulation of so-called
transAlfv\'enic turbulence (with $B_0=1$, $\delta B \sim \delta v \sim 1$);
a simulation of {\it strong} anisotropic subAlfv\'enic turbulence
(with $B_0=10$, $\delta B \sim \delta v \sim 1$).
Special attention had to be taken to the last case, where, in order for
the turbulence to be strong, the fields had to be strongly anisotropic
on the outer scale. To ensure this, the numerical box was
elongated in real space and x-direction (a direction of the strong
mean field) was 10 times longer than y and z directions. The driving
had anisotropy that corresponded to the dimensions of the box. Thus in
this case our setup is similar to one in \citet{maron2001}.

We used random solenoidal velocity driving in k-space between k=2 and 3.5.
The largest coherent eddy size L (determined by structure function technique)
was around 1/4 of a box size ($2\pi$) and the eddy turnover time for this
scale was around unity. The Alfv\'enic self-crossing time for this scale
was also around unity~\footnote{In the strong field case, $B_0=10$, the Alfv\'en speed
was ten times higher, but the largest coherent eddy was elongated with x/y aspect
ratio of around 10, so this gives the same estimate for Alfv\'enic crossing time.}.
The self-correlation timescale for our driving force was $\tau=2$ for all wavemodes.
This is somewhat longer than the eddy turnover time. We performed a test study
of the force correlation time influence on the self-correlation time of velocity,
taking $\tau=1/4,2,8$ and found no systematic dependence. The self-correlation
timescale of velocity was always around unity. We concluded that, in the case
of strong turbulence, the self-correlation timescale of velocity is primarily
determined by nonlinear interaction, rather than driving~\footnote{One can
argue that in case of the absence of nonlinear term the equation
$\partial_t {\bf v}={\bf f}$ will give infinite correlation time for $v$
even if $f$ self-correlation time is finite.}. We ran another,
higher resolution simulation, trying to find a difference in slopes between
simulations with $\tau=2$ and $\tau=6$ and found none. Note, that numerical studies
do vary in terms of $\tau$. For instance, \citet{muller2005} used driving
constant in time, and \citet{haugen2004} used $\delta$-correlated driving. Our
tentative conclusion is that the difference
in $\tau$ marginally affects the results of numerical simulations of turbulence.
In addition, in simulation M5 we used Elsasser driving~\footnote{Each of the pair
of Elsasser-field equations in MHD bear close resemblance to Navier-Stokes equation. Elsasser variables
are defined as $w^\pm=v\pm b$, where $b$ is magnetic field
in velocity units.} instead of velocity driving,
to check if it changes the tendencies observed in previous runs.

One of the advantages of driven versus decaying simulations is that it describes
a stationary random process, so, by applying ergodic hypothesis one can approximate
statistical averages with time-averages. While, in decaying simulations, if one
wants to average over time to reduce fluctuations, some hypothesis on the
decaying process has to be adopted~\footnote{Normally, one wants to normalize
the spectrum as in \citet{muller2005}, but this entails the hypothesis
that spectra are similar at different stages of decay.}.

\begin{figure}
\includegraphics[width=\columnwidth]{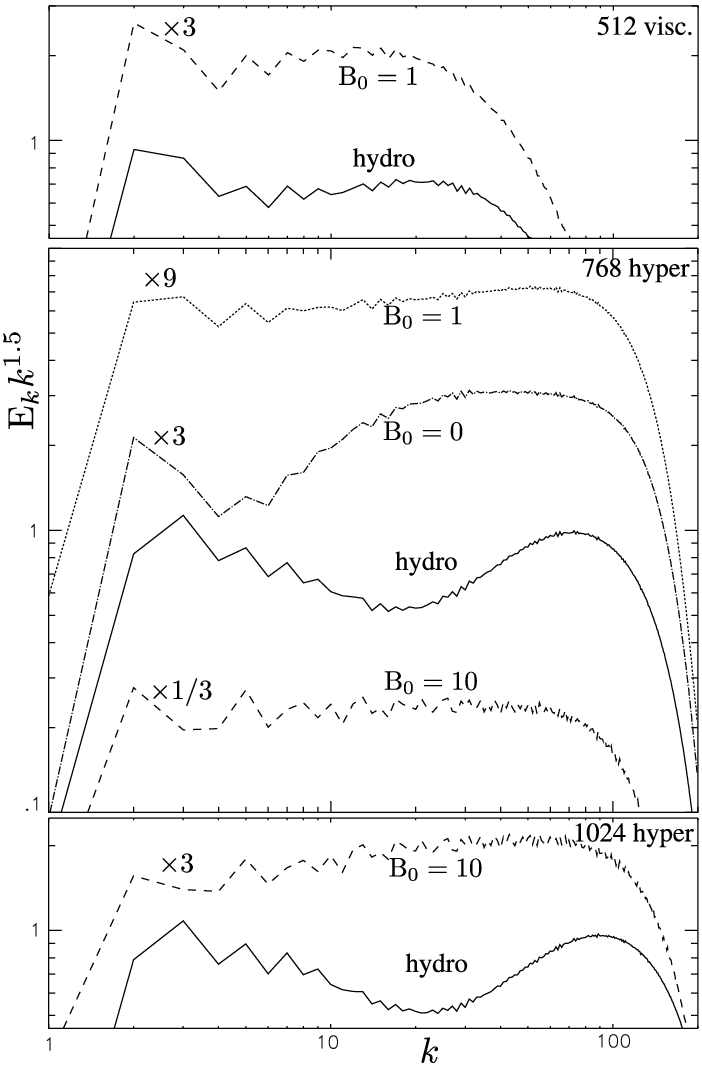}
%\plotone{spec3d.eps}
\caption{Three-dimensional angle-integrated spectra for all runs, compensated
by a factor of $k^{3/2}$. We multiplied
the spectra by a quotient of 3 to separate them from each other. Top: H1, M1;
middle: H2, M2, M3, M4; bottom: H3, M5.
A significant increase in perceived ``inertial interval'' between top plot and bottom plots
is mostly due to difference between normal viscosity and hyper-viscosity,
rather than resolution. }
\label{fig:spec3d}
\end{figure}

\section{Spectra (3D-averaged and 1D).}

Usually, bottleneck effect, a pile-up of energy near the dissipation scale,
is discussed in relation to hyper-viscosity \citep{cho2000}
or numerical viscosity \citep{kritsuk08}. But the bottleneck effect was also 
predicted \citep{falkovich} and numerically observed \citep{kaneda}
for {\it normal} viscosity. This means that the slopes, measured in
limited resolution simulations with normal viscosity, do not necessarily
reflect true asymptotic slopes.

Fig. 1 shows $E_k$, a three-dimensional power spectrum $F({\bf k})$ integrated
over angle in $k$-space:
\begin{equation}
E_k=\int_{|{\bf k}|=k}F({\bf k})d{\bf k}.
\end{equation}
This is the most popular choice for spectrum
in numerical turbulence since it is fairly straightforward to calculate.
Because $k$-space is discrete, the integration is a summation of energies
of all modes that lay in a shell of $k$-magnitudes between $n$ and $n+1$.
The number of modes, that fall into each shell, fluctuates, so this
spectrum has a characteristic toothed shape which is not remedied by
time integration. We plot $E_k$ so that our results can be directly compared
to previous numerical work that used this quantity.

In the strongly anisotropic, subAlfv\'enic simulations (as in runs
M4 M5) it makes sense to measure a perpendicular spectrum
(i.e., integrated over $k_\|$ and then over the direction of
$k_\perp$). This spectrum, however, was virtually identical to
3D spectrum integrated over solid angle, i.e., $E_k$,
so, for uniformity, we plot only $E_k$ in all cases.

The parallel spectrum (integrated
over $k_\perp$) was of no interest to us, since there are no clear
theoretical predictions for it~\footnote{One can relate this spectrum
  to the parallel structure function, calculated with respect to {\it
    global mean field}. This SF, however, do not show properties,
  predicted by GS95 model. The preferred way is to calculate SFs with
  respect to local mean field ~\citep{cho2000}.}.

\begin{figure}
\includegraphics[width=\columnwidth]{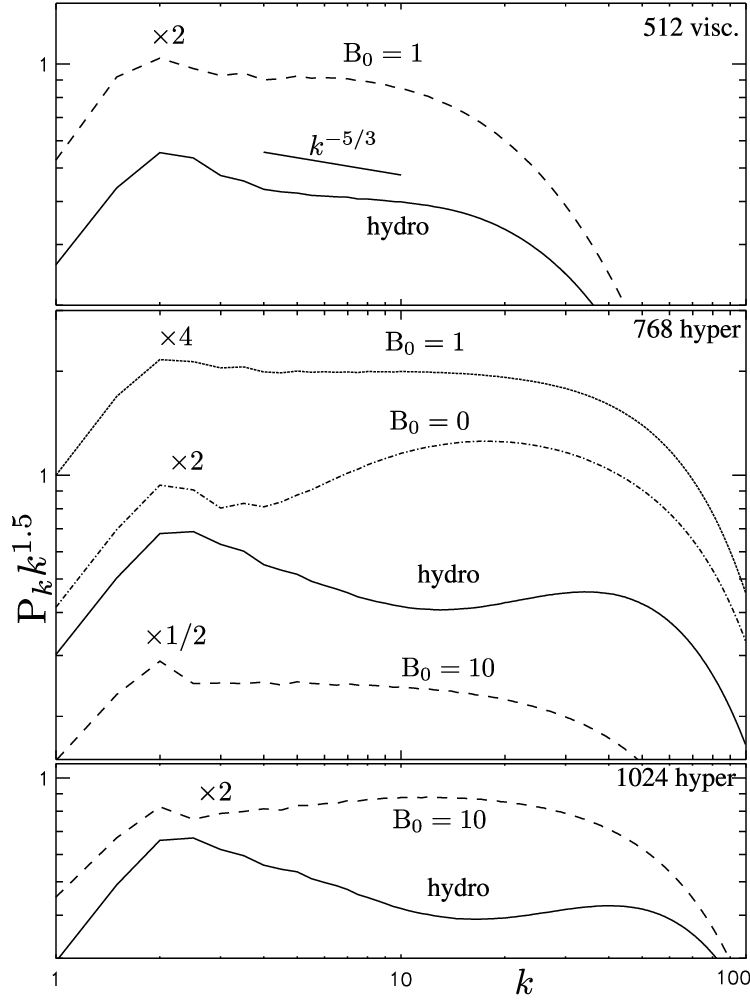}
%\plotone{spec1d.eps}
\caption{One-dimensional spectra, compensated by $k^{3/2}$.}
\label{fig:spec1d}
\end{figure}

We also measured so-called one-dimensional spectra $P_k$, which is
a power spectrum of the vector field sampled along an arbitrary line
and then averaged over ensemble. It can also be written as a Fourier
transform of the correlation function $B(r)$ \citep{monin}:
\begin{equation}
P_k=\frac1\pi\int_{-\infty}^\infty e^{-ikr}B(r)dr.
\end{equation}
Although most
simulations measure $E_k$, the structure/correlation function
scalings are the {\it primary} predictions of the Kolmogorov model, so it makes
more sense to measure $P_k$ rather than $E_k$. Also, $P_k$ is
less prone to bottleneck effect \citep{dobler2003}.
We measured $P_k$ by averaging over directions in each particular
snapshot and then averaging over time.
$P_k$ is presented in Fig 2.
For the fully statistical
homogeneous isotropic sample with infinite spacial resolution
and infinite dimensions one can derive relation $E_k=-kdP_k/dk$~\citep{monin}.
Numerically speaking, for a discrete sample with periodic boundaries
this relation is satisfied fairly well.
It ensures that in an infinite inertial
interval $E_k$ and $P_k$ will have the same slope.
In a limited resolution of our simulations
the slopes of $E_k$ and $P_k$ are fairly different though. We feel that this
difference is another useful indication that numerical determination
of slopes is rather limited (see also \S 2).

\section{Anisotropy}

\begin{figure}
\plotone{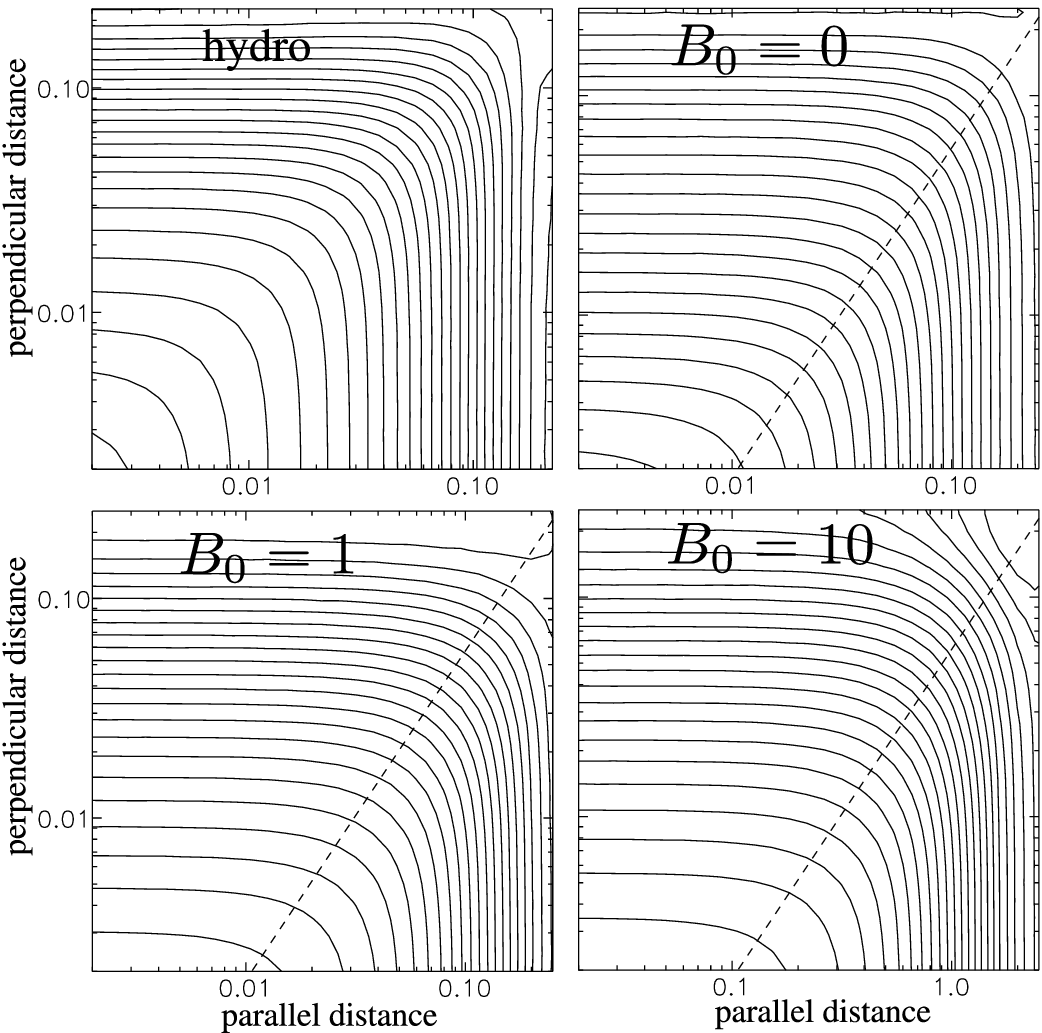}
\caption{Contours of the structure function $<|v(r+R)-v(r)|^2+|b(r+R)-b(r)|^2>_r$ where $R$ takes
arbitrary angles with respect to the local magnetic field (for hydro case this direction was chosen to be along x).
X-axis is perpendicular component, while y-axis is parallel component. Dashed line correspond to GS95 $r_\|\sim r_\perp^{2/3}$.
Hydro turbulence show isotropy on small scales, while MHD turbulence show scale-dependent anisotropy.
}
\label{fig:sf2d}
\end{figure}

Hydrodynamic turbulence is presumed to be isotropic on small scales,
while MHD turbulence is anisotropic and this anisotropy increases
to small scales without limit (GS95). The scale-dependent anisotropy,
predicted by GS95 was first observed in \citet{cho2000} by the method
of second-order structure functions and confirmed in a number of subsequent
publications. It is critical that structure function is calculated
with respect to the local magnetic field, otherwise the scale-dependent
anisotropy is not observed (see discussion on various definition
of the ``local mean field'' in Beresnyak \& Lazarian 2008).
Fig. 3 shows two-dimensional structure function where x-axis measured
distance along local mean magnetic field while y-axis measured distance
perpendicular to the field. In hydrodynamic case the ``preferred'' direction
was chosen arbitrarily. Quite predictably, the hydrodynamic turbulence
is isotropic. The MHD turbulence, however, is scale-dependently anisotropic.
For example, $B_0=0$ case is isotropic on outer scale and $B_0=1$ case is
almost isotropic on outer scale, but on small scales both are anisotropic
with anisotropy ratio of around 4. The $B_0=10$ case is anisotropic on outer
scale and this anisotropy increases by about a factor of 4 towards
small scales. These results are approximately consistent with the predictions
of GS95.

In order to study deviations from prediction of GS95, $r_\|\sim r_\perp^{2/3}$,
we created a correspondence between parallel and perpendicular scales
and plotted it on Fig. 4. This correspondence was achieved by finding
equal values of parallel and perpendicular second order structure
functions as in \citet{cho2000} or \citet{maron2001}. The observed
deviations could be connected to nonlocality,
% interaction weakening
and/or alignment effects which are discussed further.

\begin{figure}
\includegraphics[width=\columnwidth]{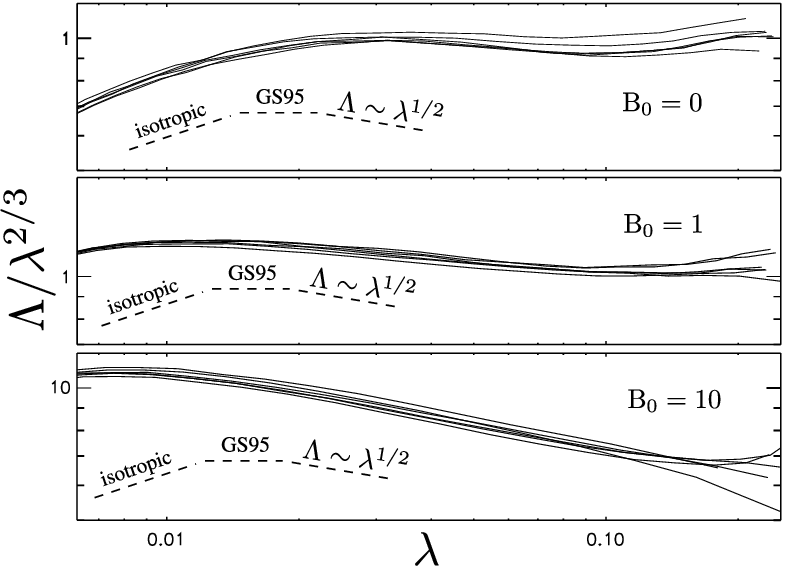}
%\plotone{lambda.eps}
\caption{Deviation of anisotropy from GS law~\cite{GS95} $\Lambda\sim \lambda^{2/3}$. M2, M3 and M4 are shown.
M5 shows behavior similar to M4, and M1 is similar to M3.
}
\label{fig:lambda}
\end{figure}

\section{Numerical evidence of interaction weakening}
In GS95 turbulent non-linear interactions are strong
and the cascading happens fast. In this
approach the Kolmogorov's $\sim -5/3$ spectral slope is expected.
In order to explain 
shallower slopes 
\citet{boldyrev2006} conjectured that interaction is depleted in
strong anisotropic turbulence by a factor which is similar to that in the
Iroshnikov-Kraichnan (IK) model, i.e. $\delta v/v_A$. In simulations
with real viscosity we obtain MHD dissipation scale $d_{\nu {\rm MHD}}$
which is somewhat larger than hydro dissipation scale
$d_{\nu {\rm HD}}$, which can be an indication of depletion of
interaction. Figs 1 and 2, upper panel, indicate that
$d_{\nu {\rm MHD}}$ is approximately 1.3 times larger than $d_{\nu {\rm HD}}$,
which is consistent with the difference in dissipation scales of
Kolmogorov and IK models, assuming the length of the inertial interval
$L/d_\nu$ of around 7. However, this result can also be explained by
the difference in Kolmogorov constant $C_K$ of hydro and MHD
turbulence. Indeed, for MHD $C_K\sim 2.2$ \citep{biskamp}, while for
hydro $C_K\sim 1.6$ \citep{frisch}, also note that $d_\nu\sim
C_K^{3/4}$. Thus, this way of proving the weakening of
interaction is still controversial.

\section{Alignment Effects.}

While most MHD turbulence models use {\it mean-field} approach and assume
that turbulence is characterized fully by spectrum or structure functions of the fields,
i.e. $\langle\delta v^2\rangle $ and $\langle\delta b^2\rangle $, recently a considerable
attention has been drawn
to so-called alignment effects~\citep{boldyrev2005,beresnyak2006,boldyrev2006}. The alignment effects can be understood, in general,
as a property of multi-variate pdf of the fields containing various
correlations. The scale-independent alignment effects are not
so interesting, because they can only modify the Kolmogorov constant of turbulence,
while scale-dependent alignment can, in principle, modify the slope.

Consider the alignment of Alfv\'en mode when all perturbations are
perpendicular to the local magnetic field, i.e. lie in the same plane.
For this purpose we use structure functions where vectors are
projected on ${\bf l\times B}$ direction where ${\bf l}$ is the direction,
connecting two points \citep{beresnyak2006}.
While a study a full multivariate PDF could be an overwhelming
task, one can introduce a few statistical measures of alignment that
could be of interest. Alignment, derived from zeroth order statistical
moments: angle alignment, 
\begin{equation}
 {\rm AA}=\langle|\sin\theta|\rangle ,
\end{equation}
 where $\theta$ is an angle between Elsasser variables perturbations
$\delta {\bf w}^+=\delta {\bf v}+\delta {\bf b}$ and
$\delta {\bf w}^-=\delta {\bf v}-\delta {\bf b}$, 
another angle alignment,
\begin{equation}
{\rm AA2}=\langle|\sin\theta_2|\rangle,
\end{equation}
where $\theta_2$ is an angle between $\delta {\bf v}$ and $\delta {\bf b}$. Alignment derived from
second order statistical moments, ``polarization intermittency''
\citep{beresnyak2006}:
\begin{equation}
{\rm PI}=\langle | \delta w^+ \delta w^- \sin \theta
|\rangle /\langle |\delta w^+ \delta w^-|\rangle;
\end{equation}
 imbalance measure
\begin{equation}
{\rm IM}=\langle |\delta (w^+)^2- \delta (w^-)^2|\rangle /\langle \delta
(w^+)^2+ \delta (w^-)^2\rangle; 
\end{equation}
imbalance correlation 
\begin{equation}
{\rm IC}=\langle|\delta w^+ \delta w^- | \rangle /\langle \delta (w^+)^2+ \delta
(w^-)^2\rangle,
\end{equation}
 velocity and magnetic fields correlation 
\begin{equation}
{\rm FC}=\langle |\delta v ||\delta
b|\rangle /\langle \delta v^2+\delta b^2\rangle 
\end{equation}
 (we found that ${\rm FC}$ is almost constant in our measurements and it will be assumed constant
thereafter). ${\rm IM}$ and ${\rm IC}$ are describing the same effect, namely
dynamic imbalance between Elsasser variables $\delta w^+$ and $\delta
w^-$, but in a different way.  For independently distributed Gaussian
fluctuations one has ${\rm AA}={\rm AA2}={\rm PI}={\rm IM}=2/\pi$, ${\rm IC}=1/\pi$.

Physically, one may have only {\it two} types of alignment,
because there are four variables (two vectors) minus normalization and
minus one arbitrary rotation along axis perpendicular to the magnetic
field. Let us choose ${\rm PI}$ and ${\rm IC}$ as two measures of alignment.
${\rm PI}$ is interesting, because it is the factor, by which an interaction
is reduced in a nonlinear
MHD term $(\delta w\cdot\nabla)\delta z \sim (\delta w\cdot k_z)\delta
z \sim \delta wk_z\delta z\sin\theta$ with respect to mean field
estimate of $\delta wk_z\delta z$ \citep{boldyrev2005}. This
quantity, along with ${\rm AA}$, was first measured in numerical simulations
by \citet{beresnyak2006}. In the subsequent publication \citet{mason2006}
used a second order structure function measure, very similar to ${\rm PI}$,
termed ``dynamic alignment'' 
\begin{equation}
{\rm DA}=\langle
|\delta v \delta b \sin \theta_2|\rangle /\langle |\delta v ||\delta
b|\rangle .
\end{equation}
It can be expressed, to a constant, as ${\rm DA}\sim {\rm PI}\cdot
{\rm IC}/{\rm FC}\sim {\rm PI}\cdot {\rm IC}$ (since ${\bf w}^+\times {\bf w}^-=-2{\bf v}\times {\bf b}$),
i.e. it is a combination of polarization
intermittency and imbalance correlation ${\rm IC}$.
It is not clear yet, whether intermittent imbalance can reduce interaction
in the balanced turbulence. For example, if one estimates
energy dissipation as $\delta w^+\delta w^-(\delta w^++\delta w^-)/\lambda$
~\citep{lithwick2007}, it is insensitive to the dynamic imbalance
$\delta w^\pm\pm\epsilon$, to the first order of
$\epsilon$. But one also may argue that the imbalanced case
is more complicated and the interaction is reduced \citep{beresnyak2008,beresnyak2009}

Numerically speaking, alignment effects were very similar for all
sub-Alfv\'enic and trans-Alfv\'enic cases\footnote{Beresnyak \& Lazarian (2006)
observed alignment not only in incompressible simulations, but
in trans-Alfvenic, trans-sonic compressible simulations as well.}.
Fig. 4 shows different alignment measures described in this section
for M5 ($B_0=10$, Elsasser driving). In the middle of
the inertial interval alignment factors depend on k as ${\rm AA}\sim
k^{0.022}$, ${\rm AA2}\sim k^{0.059}$, ${\rm PI}\sim k^{0.105}$, ${\rm IC}\sim
k^{0.084}$, ${\rm IM}\sim k^{-0.090}$, ${\rm DA}\sim k^{0.207}$.  Note, that {\rm DA} is
short of the $k^{0.25}$ dependence predicted for ``alignment angle''
in \citet{boldyrev2006}.
The alignment dependence in the velocity-driven, $B=10$ M4
simulation is somewhat different: ${\rm AA}\sim k^{0.028}$, ${\rm AA2}\sim
k^{0.052}$, ${\rm PI}\sim k^{0.137}$, ${\rm IC}\sim k^{0.054}$, ${\rm IM}\sim
k^{-0.057}$, ${\rm DA}\sim k^{0.189}$.  If one wanted to explain the
difference in slopes between M4 and M5 ($\sim 0.05$, Fig 2) by the
difference in {\rm DA} slopes (according to \citet{boldyrev2006} the ${\rm DA}$
slope $\alpha$ adds $2/3\alpha$ to the spectral slope), such an
explanation would be impossible. On the other hand, M4 and
M5 are significantly differ by ${\rm IM}$, which suggests that shallow
slopes are primarily due to imbalance.

\begin{figure}
\includegraphics[width=\columnwidth]{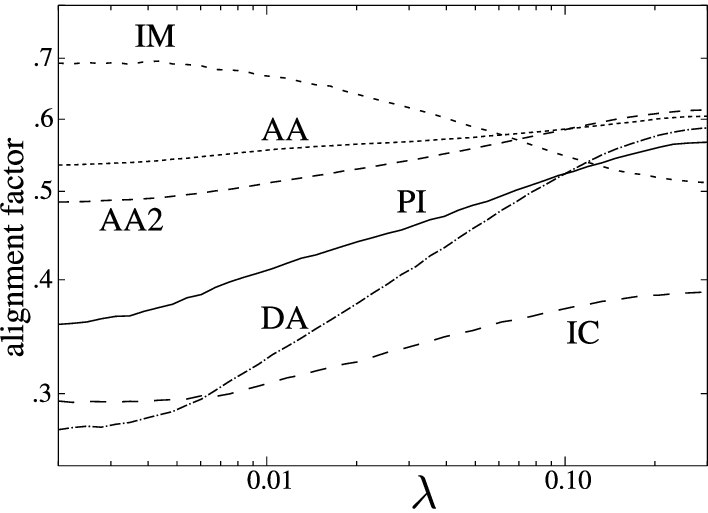}
%\plotone{align3_1.eps}
\caption{Various alignment measurement from M5.}
\label{fig:alignment}
\end{figure}

M3 ($B_0=1$, velocity driven) simulation show alignment effects
which are similar to M4. This directly contradicts
to the prediction in \citet{boldyrev2006}, that the alignment
is a function of $\delta B/B_0$, as we observe essentially
the same alignment in simulations
with $\delta B_L/B_0=1$ and $\delta B_L/B_0=0.1$ (see also Fig. 6).
We therefore conclude
that the alignment is an intermittency effect\footnote{An intermittency
correction is usually written as a function of the ration of the scale
in question to the outer scale, e.g. $\delta v_l^2=C(\epsilon l)^{2/3}(l/L)^\alpha$,
where $(l/L)^\alpha$ is the intermittency correction. In our case we also
see that the alignment is a function of $l/L$. }
 that {\it accumulates} along
the cascade, rather than being determined rigidly by $\delta B_\lambda/B_0$ as
in \citet{boldyrev2006}. Alternatively, the alignment could be a natural feature
of the nonlocal energy transfer. This calls for further investigation.

\section{Other hints on nonlocality.}
In the sub-Alfv\'enic turbulence, the properties are well-represented
by Elsasser variables $w^\pm$ and, if we assume locality of
interaction of $\delta w^+_l$ and $\delta w^-_l$, the properties of $\delta
v_l$ and $\delta b_l$ are supposed to be identical. However, we observed
notable differences even after large statistical averaging. Namely,
the magnetic energy was higher than kinetic energy in subAlfv\'enic runs
M4 and M5. The last case, which was driven by Elsasser
variables, show 50\% magnetic energy excess on large scales and around
20\% on small scales. This so-called residual energy which was proposed
to have ``-2'' spectrum scaling \citep{muller2005} but was somewhat
shallower than ``-2'' in this Elsasser driven run.
Other runs did not show any regular scaling for residual energy.
For example, statistically isotropic MHD turbulence M2
show dominance of kinetic energy on large scales (which is typical
for MHD turbulence driven with velocity on outer scale), but on small
scales magnetic energy dominates (which is typical for almost any driven
MHD turbulence). This reinforces our conjecture that currently available
3D MHD simulations do not yet exhibit inertial ranges and the flat portion
of the spectrum can not be considered a part of the inertial range
of local Kolmogorov-type turbulence until a number of other conditions
are satisfied, among which is an equipartition between spectral
kinetic and magnetic energies. 

\section{Discussion}

\subsection{Theory}
The nature of the turbulent cascade and the slope of the spectrum
of fluctuations is a central issue of turbulence and has been discussed in
a majority of papers devoted to turbulence theory and numerics.
The hydrodynamic isotropic turbulence has its ``Standard Model'' which
is based on Kolmogorov's assumption of self-similarity~\footnote{This
assumption is not precisely satisfied because of so-called intermittency.
The most popular descriptions of intermittency are probabilistic models
(Obukhov 1962, Kolmogorov 1962, She \& Leveque 1994). They give corrections to spectral slopes and higher-order
scaling exponents assumed in self-similar model}
and Kolmogorov's ``-4/5 law'':
\begin{equation}
\langle(\delta v_\|(l))^3\rangle=-\frac45\epsilon l.
\end{equation}
Assuming similarity $\langle(\delta v_\|(l))^3\rangle\sim\langle(\delta v(l))^2\rangle^{3/2}$ one
obtains the second order structure function slope of $2/3$ which correspond
to spectral slope of $-5/3$. In MHD, relations, similar to
Kolmogorov's ``-4/5 law'', exists, e.g.,
\begin{equation}
\langle\delta w_\|^\mp(l)(\delta w^\pm(l))^2\rangle=-\frac4D\epsilon^\pm l
\end{equation}
(Chandrasekhar 1951,
Politano \& Pouquet 1998)\footnote{$D=3$ for statistical isotropy and isotropic
structure function or $D=2$ for cylindrical symmetry and perpendicular structure
function, e.g. in turbulence with strong mean field.}. However, in MHD case,
this does not directly hints on scalings for $\langle(\delta w^\pm(l))^2\rangle$ and
it is not clear which similarity hypothesis has to be adopted. A nice
demonstration of this is to apply the aforementioned {\it exact} relations
to the case of {\it weak} turbulence, where, in the ansatz of three-wave
interaction, one has to obtain
$\langle\delta w_\|^\mp(l)(\delta w^\pm(l))^2\rangle\sim a(l)\langle(\delta w(l))^4\rangle/v_A$
(anisotropy $a(l)$ is defined as a ratio of parallel scale
$\Lambda(l)$ to perpendicular scale $l$),
which is very much unlike the Kolmogorov similarity hypothesis.
Thus, the spectral slope scalings are still uncertain, which stimulates
further research in this field, including attempts to measure the slope
numerically.

The nonlocality of turbulent energy transfer has been claimed in
quite a few publications. In the discussion of this numerical
work we will mention the most relevant ones and defer
more exhaustive discussion to future review papers.

It was suggested in \citet{gogoberidze} that
in MHD turbulence with strong mean field the nonlinear {\it interaction}
could be nonlocal,
with outer scale perturbations decorrelating high-frequency interacting
eddies, leading to IK-type \citep{iroshnikov,kraichnan}
interaction weakening and $-3/2$ spectrum.
However, in aforementioned model the {\it energy cascading} itself is local
\footnote{In a sense that $\epsilon=v_l^2/\tau_l$ is used, see also \S 2.}
and is performed by high-frequency eddies, i.e. there is no energy
transfer from large scales directly to small scales. We conclude that this
model can not explain the lack of bottleneck effect, observed in simulations.

In a recent model of imbalanced MHD turbulence \citep{beresnyak2008} the
eddies of the dominant component on a certain scale are aligned
{\it not} with respect to the local magnetic field on the same scale
as in \citet{cho2000} (balanced turbulence), but with respect to magnetic
field on some larger scale. One may speculate that even in the balanced case
the dynamic imbalance can cause polarization alignment, or, more likely,
that these effects are interrelated. While \citet{beresnyak2008}, by itself,
is a mean field model and reproduce locality, ``-5/3'' spectrum
and ``2/3'' anisotropy of GS95 in the balanced limit, its future
extentions to include local imbalance and polarization alignment seems promising.

\begin{figure}
\plotone{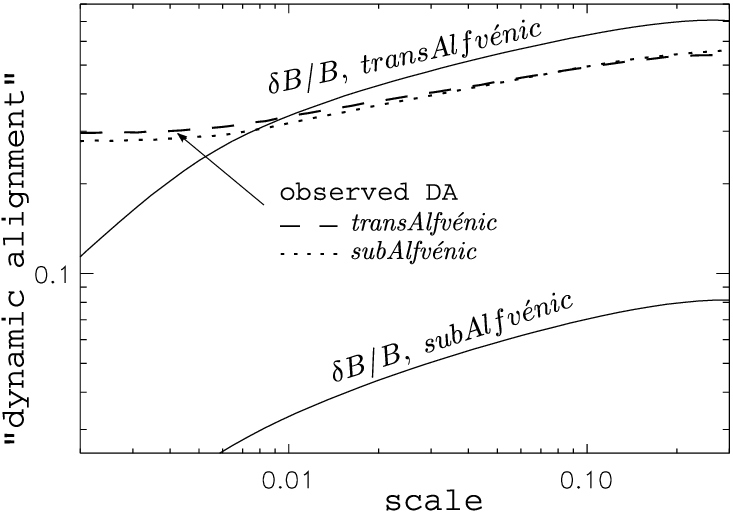}
\caption{A comparison of {\rm DA} in sub-Alfv\'enic (M4, dotted) and trans-Alfv\'enic
(MHD3b1h, dashed) simulations with $\delta B/B$ prediction of \citet{boldyrev2006}.
$\delta B/B$ is estimated from 2nd order transverse (or isotropic) structure function.}
\label{fig:compar}
\end{figure}

%new
Does the hypothesis of Boldyrev (2006) that the $v$ and $b$
alignment is limited by field wandering is justified from theory ground?
In a little thought experiment one can imagine
a perfectly aligned state where the magnitudes of $v$ and $b$ are equal.
This is a case of a perfect imbalance and also an exact solution of
MHD equations. Such a state will propagate without distortion, in other
words, no de-alignment is going to happen, although this state certainly
has some level of field wandering. This thought experiment hints that
the hypothesis of $v$ and $b$ alignment being limited by field wandering
directly contradicts MHD equations. The effect
of local imbalance, therefore, has to be treated in a more complicated way.
Fig. ~\ref{fig:compar} shows a comparison between field wandering ($\delta B/B$)
and ``dynamic alignment'' in transAlfv\'enic M3 and subAlfv\'enic M4.
We see that in subAlfv\'enic case the dynamic alignment is an order of magnitude
off the $\delta B/B$, which is in contrast with \citet{boldyrev2006} which suggest
that ${\rm DA}\sim \delta B/B$.

\subsection{Previous numerical work}
The study of weakly compressible MHD turbulence by \citet{haugen2003,haugen2004}
revealed some difference in bottleneck effect in hydro and MHD cases, although
the authors reported bottleneck behaviour as ``very similar''. The difference
was small, possibly, due to adopted first order viscous and resistive dissipation.  
Aforementioned work, however, used finite-difference code and, inadvertently,
the numerical dissipation had a different character in MHD and hydro
simulations, which precluded a rigorous comparison. Nevertheless,
we can say that it is consistent with what we see in our incompressible simulations.

Biskamp et al. (1998) studied two-dimensional MHD and EMHD turbulence.
They noticed that both MHD and EMHD cases have an unusual nonlocal
bottleneck effect, which appeared differently depending on numerical
resolution. It is possible that spectral flattening from such an
effect can be perceived as a ''false'' inertial interval and lead
to an incorrect estimate of the slope. Although there had been
much discussion on the analogy between two-dimensional MHD turbulence
and MHD turbulence with strong mean field, the predominant picture
is that Alfv\'enic turbulence is essentially three-dimensional (GS95, Cho \& Vishniac 2000,
Maron \& Goldreich 2001, Cho et al. 2002).

Yousef et al. (2007) claimed that in the statistically isotropic
MHD turbulence (similar to our simulation M2) one
has a folded magnetic field structures that directly non-locally
interact with outer-scale motions. They concluded that one can not
use Kolmogorov argumentation because of this nonlocality.
We, however, belive, that Kraichnan's argumentation 
regarding the dominance of local mean field, i.e. the
Alfv\'en effect, is correct even in the case of $B_0=0$.
This is somewhat hard to demonstrate in 3D numerics, however,
because for $B_0=0$ there is a transition region between
outer scale and inertial interval 
where kinetic energy dominates and the magnetic spectrum is
very flat, i.e., there is no clear dominance of the large-scale
magnetic field. Also the use of first-order (natural) viscosity
in aforementioned three-dimensional simulations made the
inertial range very short, which created
an illusion of a universal folded magnetic field structure.

Alexakis et al. (2005a, 2005b)
used a specific numerical tool to quantify the transfer
of energy between scales in both hydro and MHD turbulence. They
claimed that MHD case is somewhat nonlocal. However, 
the more radical claim is that the hydro cascade
and the larger part of MHD cascade
is extremely local, i.e. the energy is not transferred
between k and 2k, as in Kolmogorov model, but
instead between k and $k+k_0$ where $k_0$ is
determined by outer scale, which breaks
self-similarity. We find these results rather
puzzling and defer discussion until independent
confirmation is available. Until then, we assume
that Kolmogorov picture is roughly appropriate
for isotropic hydrodynamic turbulence with large
inertial range.

\subsection{Implications for astrophysical turbulence}
Realistic astrophysical turbulence is, in general, compressible.
The examples of weakly compressible flows are the quiet convection
in main sequence stars and turbulence in very hot intracluster gas.
The turbulence in most of the Interstellar Medium,
however, is strongly compressible, due to effective cooling.
This raises the question of to what extend
the results of incompressible simulations are applicable to astrophysics.
Intuitively, one could expect that weak small-amplitude Alfv\'en and slow-mode
turbulent perturbations should be nearly incompressible.
The question is how to deal with the fast mode and large-amplitude MHD
turbulence in compressible fluids.
% and also of the application of the strong MHD turbulence theory, if the
%perturbations are weak. The answer to the latter issue is related to the limited inertial range of the 
%weak turbulence. Indeed, it is possible to show that if the turbulence is injected at the scale $l$ with 
%the velocity $v_l<V_A$, the inertial range of the weak turbulence will be $[l, l(v_l/V_A)^2]$ and
% for scales less than $l(v_l/V_A)^2$ the turbulence is strong (see Lazarian \& Vishniac 1999).
% The corresponding velocities will be  subsonic and nearly incompressible
% provided that $v_l(v_l/V_A)^2\ll V_s$, where $V_s$ is the sound velocity.
%The other issue, namely, 

The coupling of Alfv\'enic motions and compressible motions is a difficult subject.
Theoretical arguments that fast, slow and Alfv\'en modes may create independent energy
cascades were provided in GS95, Lithwick \& Goldreich (2001) and Cho \& Lazarian (2003).
These arguments were applicable to strong Alfv\'enic turbulence\footnote{The more 
detailed study (Chandran 2005) of {\it weak} Alfv\'enic turbulence in a pressure-less medium
showed that the interactions between the fast and Alfv\'enic modes may be significant in a certain
regions of k-space.}.
Cho \& Lazarian (2002, 2003) numerically demonstrated that for even for appreciable Mach numbers (up to 10),
the properties of Alfv\'enic modes (scalings and anisotropy) were similar to those in incompressible simulations\footnote{Density perturbations, which are absent in incompressible flows,
are definitely affected by compressible motions. However, Beresnyak et al. (2005)
found that the structure of logarithm of density reveal scale-dependent anisotropy, similar
to GS95 law. This presumes that a significant fraction
of density structures are created by shearing by Alfv\'enic perturbations.}.
% When actual densities are concerned, the latter
%result is only valid for subsonic driving (Cho \& Lazarian 2003).}.
They also showed that the decay time for the Alfv\'enic modes is fast and not mediated
through coupling with compressible motions, which used to be the common wisdom at the time of the study.
The effect of scale-dependent polarization alignment, a characteristic of Alfv\'enic
cascade discussed in this paper, happen to be present in both compressible and incompressible
MHD simulations (Beresnyak \& Lazarian 2006).
While the extend to which strong shocks can modify the Alfv\'enic cascade deserves more study,
the arguments above make us confident that the studies of incompressible turbulence
are of primary importance to understand astrophysical turbulence.

There is another issue, which is frequently ignored when one compares numerical MHD with interstellar
turbulence. If magnetic fields are perfectly frozen into the fluid, turbulence creates multiple
small-scale current sheets which are difficult to dissipate.
Within such zones the frozen-in condition is no longer valid
and magnetic reconnection takes place (see Biskamp 2000, Priest \& Forbes 2000, Bhattacharjee 2004, Zweibel \& Yamada 2009). The Lundquist number $S\equiv LV_A/\eta$ that characterizes how well magnetic fields are frozen in ($L$ is the scale of the current sheet and $\eta$ is magnetic diffusivity), is very high,
e.g. $>10^{10}$, for most astrophysical fluids and is fairly low, e.g. $<10^{4}$, for MHD simulations.
If magnetic reconnection in astrophysics depends on $S$, this presents not only a problem for most of MHD simulations, e.g. simulations of dynamo, molecular clouds, accretion disks, but also means that the numerical results on turbulent scalings may not be trusted.
We believe that the extensive observational data suggests that magnetic reconnection is fast. Also,
a model predicting fast reconnection in turbulent fluids, Lazarian \& Vishniac (1999),
has been successfully tested in Kowal et al. (2009). This gives additional support to
numerical testing of astrophysical turbulence.

The slopes and turbulence anisotropies are important for a variety of astrophysical phenomena. For instance, scattering and turbulent acceleration of cosmic rays depends on the scaling of MHD turbulence (see Chandran 2000, Yan \& Lazarian 2002, 2004). So does the perpendicular diffusion of cosmic rays and heat transport in plasmas (Narayan \& Medvedev 2001, Lazarian 2006).
Naturally, it is important to establish the true scalings of MHD turbulence. This paper testifies
that a higher resolution numerical simulations are required for accurate testing
of the present and future MHD turbulence models.

\section{Conclusions}
We conclude that although the Kolmogorov-like
Goldreich-Sridhar (GS95) model is appealing, simple
and captures some essential physical properties of the strong
MHD turbulence, such as scale-dependent anisotropy,
it should be amended to explain cascade nonlocality
and scale-dependent alignment effects.
How to achieve this is the issue of the 
future research, as we demonstrate that the existing attempts to improve GS95
do not agree well with the presented numerical simulations. In addition, we issue a note
of warning that the numerical measurements of the spectral slope that served as a 
motivation for many of theoretical studies are unlikely to represent the true 
theoretical slopes due to the non-locality of MHD cascade.

\acknowledgments
%\begin{acknowledgments}
AB thanks IceCube project for support of his research.
AB thanks TeraGrid and NCSA for providing computational resources.
AL acknowledges the NSF grant AST-0808118 and support from
the Center for Magnetic Self-Organization.
Both authors are grateful to anonymous referee for useful suggestions.
%\end{acknowledgments}

\end{document}